\documentclass[11pt]{article}
\usepackage{type1cm}
\usepackage[dvipdfmx]{graphicx}
\usepackage{cite}

\usepackage{amsmath,amssymb,bm}

\def\beq{\begin{equation}}
\def\eeq{\end{equation}}
\def\nbeq{\begin{equation*}}
\def\neeq{\end{equation*}}
\def\<{\langle}
\def\>{\rangle}

\renewcommand{\d}{\partial}

\begin{document}
\title{Quasi-equilibrium nonadditivity}
\author{Takashi Mori \\
{\it
Department of Physics, Graduate School of Science,} \\
{\it The University of Tokyo, Bunkyo-ku, Tokyo 113-0033, Japan
}}
\maketitle

\begin{abstract}
The possibility that a short-range interacting system exhibits nonadditivity is investigated.
After the discussion on the precise definition of additivity and its consequence,
we show that it is possible when the system is in a quasi-equilibrium state by examining the specific model, in which the spin degrees of freedom are coupled to the motional degrees of freedom
and which exhibits a type of structural phase transitions.
\end{abstract}

\section{Introduction}

In short-range interacting systems, in which the interaction potential satisfies suitable conditions, 
it is rigorously proven that the thermodynamic limit exists and the microcanonical entropy is concave~\cite{Ruelle_text}.
As a result of concavity of the entropy, some desirable properties consistent with thermodynamics immediately follow.
For instance, the microcanonical ensemble is equivalent to the canonical ensemble
and the entropy is connected to the free energy via the Legendre transformation.
One of the important consequences of the ensemble equivalence is that the specific heat and other susceptibilities are always non-negative.

In long-range interacting systems, on the other hand, nontrivial thermodynamic limit exists if Kac's prescription is applied~\cite{Campa_review2009,Levin_review2014},
but the limiting microcanonical entropy is no longer concave.
Thus the ensemble equivalence can be violated and the specific heat in the microcanonical ensemble may be negative~\cite{Thirring1970}.
Such anomalous behavior is understood to be the result of lack of additivity.
Additivity means that a macroscopic system can be divided into the two subsystems without a macroscopic amount of work,
which is, sometimes implicitly, assumed in usual thermodynamics.
In long-range interacting systems, the additivity is not satisfied due to the strong coupling between the two subsystems.

In this paper,we will see that a {\it short}-range interacting system can be nonadditive and behave like a long-range interacting system 
as long as the system is in a {\it quasi-equilibrium state}.
Some systems do not thermalize directly but first reaches a long-lived metastable state, and then relax to the genuine thermal equilibrium.
If this metastable state is described by equilibrium statistical mechanics of some effective Hamiltonian, it is referred to as a quasi-equilibrium state,
see  Sec.~\ref{sec:QES} for a precise definition of quasi-equilibrium states.
Depending on the given experimental timescale, there is a situation in which what we actually observe is not a true equilibrium but a quasi-equilibrium state.
Rigorous results of statistical mechanics~\cite{Ruelle_text} do not cover quasi-equilibrium states, 
and thus they do not exclude the possibility that the ensemble equivalence is violated and the specific heat is negative even in a short-range interacting system
when it is in a quasi-equilibrium state.

In equilibrium statistical mechanics, the concept of time does not appear in the theory,
which reflects the separation of timescales.
Feynman characterized the concept of thermal equilibrium as follows~\cite{Feynman_text}:
``if all the ``fast'' things have happened and all the ``slow'' things not, the system is said to be in {\it thermal equilibrium}''.
If there are several distinct timescales, correspondingly there will be several effective Hamiltonians.
When we consider quasi-equilibrium phenomena for a specific timescale,
we must choose a proper effective Hamiltonian and study its equilibrium statistical mechanics.
Statistical mechanics of the ``exact Hamiltonian'' does not necessarily describe the ``equilibrium'' state which is actually observed.
From the fundamental point of view, it is important to understand the possible behavior of macroscopic systems in quasi-equilibrium states.

The outline of this paper is the following.
In Sec.~\ref{sec:setup}, theoretical setup is explained.
Then we discuss the fundamental concepts, additivity in Sec.~\ref{sec:additivity} and quasi-equilibrium states in Sec.~\ref{sec:QES}.
After that we explain the theoretical model and discuss its properties in Sec.~\ref{sec:model}.
In Sec.~\ref{sec:properties}, the detailed properties of the quasi-equilibrium states are examined.
We will see that the model displays nonadditivity in quasi-equilibrium states.
In Sec.~\ref{sec:effective}, we show that the elastic motion mediates the effective long-range spin-spin interactions in quasi-equilibrium states.
Finally, we conclude the paper in Sec.~\ref{sec:conclusion}.

\section{General setup}
\label{sec:setup}

We consider a classical system composed of $N$ identical particles in the domain denoted by $\Gamma\subset\mathbb{R}^d$.
The volume of the system is denoted by $V=|\Gamma|$, and we use the notation such as $l\Gamma=\{\bm{x}\in\mathbb{R}^d:\bm{x}/l\in\Gamma\}$
and $\Gamma+\bm{a}=\{\bm{x}\in\mathbb{R}^d:\bm{x}-\bm{a}\in\Gamma\}$ for $l\in\mathbb{R}$ and $\bm{a}\in\mathbb{R}^d$.
A microscopic state of $i$th particle is denoted by $\xi_i\in{\cal S}$, and we define $\bm{\xi}=(\xi_1,\xi_2,\dots,\xi_N)\in {\cal S}^N$.

In an isolated system of volume $V=|\Gamma|$, thermodynamic properties are characterized by the entropy $S$, whose density is defined as
\beq
s(\varepsilon,m;\Gamma)=\frac{S}{V}=\frac{1}{V}\ln\sum_{\bm{\xi}\in{\cal S}^N}
\delta(Vm\leq M(\bm{\xi})\leq Vm+\Delta M)\delta(V\varepsilon\leq H(\bm{\xi})\leq V\varepsilon+\Delta E),
\label{eq:entropy}
\eeq
where $\varepsilon$ is the energy density and
\beq
\delta(A)=\begin{cases}1 & \text{when $A$ is true}
\\ 0 & \text{otherwise}.
\end{cases}
\eeq
Here, $M(\bm{\xi})$ denotes a set of {\it completely additive} thermodynamic quantities~\cite{Touchette2002} 
such as the number of particles and the magnetization.
$M(\bm{\xi})$ is said to be completely additive if it is expressed as
\beq
M(\bm{\xi})=\sum_{i=1}^NM(\xi_i).
\eeq
When $\xi\in{\cal S}$ takes continuous values, the summation over $\bm{\xi}$ is replaced by the corresponding integral.
As usual, $\Delta E$ and $\Delta M$ in Eq.~(\ref{eq:entropy}) are chosen so that they are large enough to contain many microscopic states in the given interval but macroscopically very small.

A system in contact with a thermal reservoir at a temperature $T=\beta^{-1}$, where Boltzmann's constant is set to unity throughout this paper,
is described by the canonical ensemble.
The corresponding thermodynamic potential is the free energy $F$, whose density is given by
\beq
f(\beta,m;\Gamma)=\frac{F}{V}=-\frac{1}{V\beta}\ln\sum_{\bm{\xi}\in{\cal S}^N}\delta(Nm\leq M(\bm{\xi})\leq Nm+\Delta M)\exp[-\beta H(\bm{\xi})].
\eeq
When $\xi$ is not conserved and instead the external field $h$ conjugate to $M$ is applied, the free energy density is given as a function of $\beta$ and $h$ by
\beq
f(\beta,h;\Gamma)=-\frac{1}{V\beta}\ln\sum_{\bm{\xi}\in{\cal S}^N}\exp[-\beta(H(\bm{\xi})-hM(\bm{\xi}))].
\eeq

For a lattice system, it is sometimes convenient to consider the specific entropy and the specific free energy, 
$S/N$ and $F/N$, rather than $S/V$ and $F/V$.
We will use the same notation $s(\varepsilon,m;\Gamma)$, $f(\beta,m;\Gamma)$, and $f(\beta,h;\Gamma)$ 
for the quantities per particle in Sec.~\ref{sec:properties}.

In this paper, the thermodynamic limit is taken in such a way that $L\rightarrow\infty$ with $\Gamma=L\gamma$,
where $\gamma\subset\mathbb{R}^d$ is some finite domain.
That is, we make the system large without changing its shape~\footnote
{In usual short-range interacting systems, the entropy density is independent of the shape of the system~\cite{Ruelle_text}.
However, in long-range interacting systems with the appropriate scaling called Kac's prescription, 
the entropy density in the thermodynamic limit exists but depends on $\gamma$.}.
The entropy density in the thermodynamic limit is denoted by
\beq
s_{\gamma}(\varepsilon,m)=\lim_{L\rightarrow\infty}s(\varepsilon,m;L\gamma).
\eeq
Correspondingly, the thermodynamic limit of the free energy density is denoted by $f_{\gamma}(\beta,m)$ or $f_{\gamma}(\beta,h)$ depending on the situation.

\section{Definition of additivity}
\label{sec:additivity}

We give a precise definition of additivity in this section because the term of additivity has been sometimes used vaguely,
see also Ref.~\cite{Touchette2002} for similar but another definition of additivity.

In this paper, we define additivity from the thermodynamic point of view.
That is, {\it if we can divide a macroscopic system into the two subsystems, say $A$ and $B$, by a quasi-static adiabatic process without a macroscopic amount of work, the system is said to be additive}.

Starting from this thermodynamic definition of additivity, we shall discuss its meaning in the statistical-mechanical point of view.
The Hamiltonian $H(\bm{\xi})$ is suitably divided into the three parts as
\beq
H(\bm{\xi})=H_A(\bm{\xi}_A)+H_B(\bm{\xi}_B)+H_{\rm int}(\bm{\xi}_A,\bm{\xi}_B),
\eeq
where $\bm{\xi}_A=\{\xi_i\}_{i\in A}$ and $\bm{\xi}_B=\{\xi_i\}_{i\in B}$.
If there is only the subsystem $A$ or $B$, its Hamiltonian is $H_A$ or $H_B$, respectively.

Let
\beq
H_{\eta}(\bm{\xi})=H_A(\bm{\xi}_A)+H_B(\bm{\xi}_B)+\eta H_{\rm int}(\bm{\xi}_A,\bm{\xi}_B).
\eeq
Then a quasi-static adiabatic process to divide the system into $A$ and $B$ is described by changing $\eta$ very slowly from 1 to 0.
Now we consider the case in which each of the completely additive quantities of $A$ and $B$ is held fixed during the process.
We denote those densities by $m_A$ and $m_B$.
When the system is isolated from the environment, the amount of work done by the system during this thermodynamic process is given by
\beq
W_{\rm adiabatic}=V(\varepsilon-\varepsilon'),
\label{eq:micro_work}
\eeq
where $\varepsilon$ and $\varepsilon'$ are the energy densities before and after the thermodynamic process, respectively.
Since the entropy is invariant under the quasi-static adiabatic process, $\varepsilon'$ is determined by the condition
\beq
\left. s(\varepsilon,m_A,m_B;\Gamma_A,\Gamma_B)\right|_{\eta=1}=\left. s(\varepsilon',m_A,m_B;\Gamma_A,\Gamma_B)\right|_{\eta=0},
\label{eq:adiabatic}
\eeq
where $\Gamma_A\subset\mathbb{R}^d$ and $\Gamma_B\subset\mathbb{R}^d$ are the domains of the subsystems $A$ and $B$, respectively, and they satisfy $\Gamma_A\cap\Gamma_B=\emptyset$ and $|\Gamma_A\cup\Gamma_B|=V$.
The function $s(\varepsilon,m_A,m_B;\Gamma_A,\Gamma_B)$ is the entropy density of the composite system $A$ and $B$
under the condition that both the values of $m_A$ and $m_B$ are held fixed but $A$ and $B$ can exchange the energy with each other.
If
\beq
\lim_{V\rightarrow\infty}\frac{W_{\rm adiabatic}}{V}=\lim_{V\rightarrow\infty}(\varepsilon-\varepsilon')=0,
\label{eq:additivity_micro}
\eeq
the system is said to be additive. 
Here the limit of $V\rightarrow\infty$ means the limit of $L\rightarrow\infty$ with $\Gamma_A=L\gamma_A$ and $\Gamma_B=L\gamma_B$.
This is the precise definition of additivity.

The entropy densities associated with $H_A$ and $H_B$ are given by 
$s(\varepsilon_A,m_A;\Gamma_A)$ and $s(\varepsilon_B,m_B;\Gamma_B)$, respectively.
Equations~(\ref{eq:adiabatic}) and (\ref{eq:additivity_micro}) then yield
\beq
s(\varepsilon,m_A,m_B;\Gamma_A,\Gamma_B)
=\sup_{\substack{\varepsilon_A,\varepsilon_B: \\ \lambda\varepsilon_A+(1-\lambda)\varepsilon_B=\varepsilon}}
\left[\lambda s(\varepsilon_A,m_A;\Gamma_A)+(1-\lambda)s(\varepsilon_B,m_B;\Gamma_B)\right]+o(1)
\eeq
with $\lambda=|\Gamma_A|/|\Gamma_A\cup\Gamma_B|$.
Now we take the limit of $L\rightarrow\infty$.
In this thermodynamic limit, the entropy for $\gamma_A\cup\gamma_B$ is obtained by
maximizing the sum of the entropies for $\gamma_A$ and $\gamma_B$ with respect to the distribution of $m_A$ and $m_B$.
That is,
\beq
s_{\gamma_A\cup\gamma_B}(\varepsilon,m)
=\sup_{\substack{m_A,m_B: \\ \lambda m_A+(1-\lambda)m_B=m}}s_{\gamma_A,\gamma_B}(\varepsilon,m_A,m_B),
\label{eq:entropy_contraction}
\eeq
where
\beq
s_{\gamma_A,\gamma_B}(\varepsilon,m_A,m_B)=\lim_{L\rightarrow\infty}s(\varepsilon,m_A,m_B;L\gamma_A,L\gamma_B).
\eeq
In this way, the condition of additivity, Eq.~(\ref{eq:additivity_micro}) is rewritten as
\beq
s_{\gamma_A,\gamma_B}(\varepsilon,m_A,m_B)
=\sup_{\substack{\varepsilon_A,\varepsilon_B: \\ \lambda\varepsilon_A+(1-\lambda)\varepsilon_B=\varepsilon}}
\left[\lambda s_{\gamma_A}(\varepsilon_A,m_A)+(1-\lambda)s_{\gamma_B}(\varepsilon_B,m_B)\right]
\label{eq:entropy_additivity}
\eeq
for any pair of $m_A$ and $m_B$.
This is another expression of the thermodynamic definition of additivity.
In a short-range interacting system satisfying the suitable conditions~\cite{Ruelle_text},
it can be rigorously shown that Eq.~(\ref{eq:entropy_additivity}) holds.

Equations~(\ref{eq:entropy_contraction}) and (\ref{eq:entropy_additivity}) imply
\beq
s_{\gamma_A\cup\gamma_B}(\varepsilon,m)=\sup_{\substack{\varepsilon_A,\varepsilon_B: \\ \lambda\varepsilon_A+(1-\lambda)\varepsilon_B=\varepsilon}}
\sup_{\substack{m_A,m_B: \\ \lambda m_A+(1-\lambda)m_B=m}}
\left[\lambda s_{\gamma_A}(\varepsilon_A,m_A)+(1-\lambda)s_{\gamma_B}(\varepsilon_B,m_B)\right].
\label{eq:additivity_contraction}
\eeq
Now the following fact immediately follows: {\it the additivity implies that the entropy density is independent of the shape of the system}.
This conclusion is obtained by using Eq.~(\ref{eq:additivity_contraction}) and assuming the translational symmetry of the Hamiltonian, 
which yields $s_{\gamma}(\varepsilon,m)=s_{\gamma+\bm{a}}(\varepsilon,m)$ for an arbitrary $\bm{a}\in\mathbb{R}^d$.
Therefore, we can simply put $s_{\gamma}(\varepsilon,m)=s(\varepsilon,m)$.
By using this property, it is also obvious that 
$s_{\gamma_A,\gamma_B}(\varepsilon,m_A,m_B)$ only depends on $\lambda=|\gamma_A|/|\gamma_A\cup\gamma_B|$.
Therefore, it will not be confusing to write $s_{\gamma_A,\gamma_B}(\varepsilon,m_A,m_B)=s(\varepsilon,m_A,m_B)$.

As a result of the above fact, Eq.~(\ref{eq:additivity_contraction}) is equivalent to the concavity of the entropy,
\beq
s(\lambda\varepsilon_A+(1-\lambda)\varepsilon_B,\lambda m_A+(1-\lambda)m_B)\geq \lambda s(\varepsilon_A,m_A)+(1-\lambda)s(\varepsilon_B,m_B).
\eeq
Therefore, {\it the entropy of an additive system is always concave}~\footnote
{The converse is not true in general.
In some systems, there is a situation in which the entropy is concave but the system is not additive.
Such examples include the spin-1/2 quantum $XXZ$ model with infinite-range interactions~\cite{Mori_equilibrium2012,Kastner_statmech2010}.}.
Equivalently, the system is not additive if the entropy is not a concave function of $\varepsilon$ and $m$.
Practically we can judge whether the system is additive or not by measuring $W_{\rm adiabatic}$.

The concavity of the entropy as a function of the energy density and the densities of completely additive quantities is a very important property in statistical mechanics.
It ensures equivalence of several statistical ensembles, e.g. the microcanonical, canonical, and grandcanonical ensembles.
In Ref.~\cite{Touchette_arXiv2014}, a detailed account on the deep relations between the concavity of the entropy and the ensemble equivalence of several levels can be found.

When the system is in contact with a thermal reservoir, the amount of work done by the system during the quasi-static isothermal process
of changing $\eta$ from 1 to 0 with each of $m_A$ and $m_B$ held fixed is
\beq
W_{\rm isothermal}=V\left[\left. f(\beta,m_A,m_B;\Gamma_A,\Gamma_B)\right|_{\eta=1}
-\left. f(\beta,m_A,m_B;\Gamma_A,\Gamma_B)\right|_{\eta=0}\right].
\eeq
We denote
\beq
f(\beta,m_A,m_B;\Gamma_A,\Gamma_B)|_{\eta=1}=f(\beta,m_A,m_B;\Gamma_A,\Gamma_B)
\eeq
and
\beq
f(\beta,m_A,m_B;\Gamma_A,\Gamma_B)|_{\eta=0}=\lambda f(\beta,m_A;\Gamma_A)+(1-\lambda)f(\beta,m_B;\Gamma_B).
\eeq
Again putting $\lambda=|\Gamma_A|/|\Gamma_A\cup\Gamma_B|=|\gamma_A|/|\gamma_A\cup\gamma_B|$
and taking the limit of $L$ going to infinity with $\Gamma_A=L\gamma_A$ and $\Gamma_B=L\gamma_B$,
we can show that the condition
\beq
\lim_{V\rightarrow\infty}\frac{W_{\rm isothermal}}{V}=0,
\label{eq:additivity_can}
\eeq
is equivalent to
\beq
f_{\gamma_A,\gamma_B}(\beta,m_A,m_B)=\lambda f_{\gamma_A}(\beta,m_A)+(1-\lambda)f_{\gamma_B}(\beta,m_B).
\label{eq:free_additivity}
\eeq
From this equation, the translational symmetry of the Hamiltonian, and
\beq
f_{\gamma_A\cup\gamma_B}(\beta,m)=\inf_{\substack{m_A,m_B: \\ \lambda m_A+(1-\lambda)m_B=m}}f_{\gamma_A,\gamma_B}(\beta,m_A,m_B),
\eeq
it follows that $f_{\gamma}(\beta,m)$ is independent of $\gamma$ and it is simply denoted by $f(\beta,m)$.
Similarly, $f_{\gamma_A,\gamma_B}(\beta,m_A,m_B)$ only depends on $\lambda$, and it will not be confusing 
if we simply put $f_{\gamma_A,\gamma_B}(\beta,m_A,m_B)=f(\beta,m_A,m_B)$.

We can also derive Eq.~(\ref{eq:free_additivity}) from Eq.~(\ref{eq:entropy_additivity}) by using the relation
\beq
f(\beta,m)=\inf_{\varepsilon}\left[\varepsilon-\frac{1}{\beta}s(\varepsilon,m)\right],
\eeq
which is always correct as long as the thermodynamic limit exists.
In general, however, Eq.~(\ref{eq:entropy_additivity}) cannot be derived from Eq.~(\ref{eq:free_additivity}).
If Eq.~(\ref{eq:free_additivity}) does {\it not} hold, Eq.~(\ref{eq:entropy_additivity}) also does not hold and the system is {\it not} additive.

Equations~(\ref{eq:entropy_additivity}) and (\ref{eq:free_additivity}) can be seen as the statistical-mechanical expressions of the additivity.
In order to explain this aspect, let us consider an additive system in a thermal reservoir.
If a completely additive quantity $M(\bm{\xi})$ is not conserved and fluctuates,
the probability $P(m)dm$ of $M(\bm{\xi})/V$ lying between $m$ and $m+dm$ is proportional to $\exp[-\beta Vf(\beta,m)]dm$, which is known as Einstein's formula~\cite{Landau_stat}.
Therefore, Eq.~(\ref{eq:free_additivity}) corresponds to the following relation:
\beq
P_{AB}(m_A,m_B)\sim P_A(m_A)P_B(m_B),
\eeq
where $P_{AB}(m_A,m_B)$ is the joint probability distribution of $m_A$ and $m_B$ in the composite system.
In this way, the thermodynamic definition of additivity corresponds to the statistical independence of the two macroscopic subsystems.

\subsection*{Remark on another conventional definition of additivity}
In some literatures, additivity is defined by the smallness of $H_{\rm int}$.
If $H_{\rm int}$ is small, the two subsystems $A$ and $B$ will be almost independent of each other.
In a short-range interacting system, the value of $H_{\rm int}$ is typically proportional to the surface area between the two subsystems, and hence it is very small compared to $H_A$ and $H_B$, both of which are proportional to the volume of the system.
On the other hand, in a long-range interacting system, $H_{\rm int}$ can become very large and additivity can be violated.

For many cases this naive definition works well, but, in principle, there may be some situations in which the two subsystems are not independent of each other although $H_{\rm int}$ is very small.
That is, if the interaction itself is short-ranged but its influence spreads out over long distances,
the two subsystems cannot be regarded as independent
although $H_{\rm int}$ is proportional to the surface area and therefore small.
In Sec.~\ref{sec:Hamiltonian}, we will consider such a situation.

In this way, to define additivity by the smallness of $H_{\rm int}$ is too naive.
On the other hand, Eq.~(\ref{eq:additivity_micro}) can be interpreted as the {\it smallness of the influence} of $H_{\rm int}$.

\section{Quasi-equilibrium state}
\label{sec:QES}

The (micro)canonical ensemble for the Hamiltonian $H(\bm{\xi})$ describes an equilibrium state of the system whose energy is given by $H(\bm{\xi})$.
An equilibrium state is realized after the time evolution from an initial state for a sufficiently long time.
However, in some systems with certain initial conditions, the system is first relaxed to a metastable state,
and after a very long time it eventually reaches an equilibrium state.
If the lifetime of this metastable state is much longer than any experimental timescale,
what we observe as a steady state is not a true equilibrium but a metastable state.

If a metastable state is described by the (micro)canonical ensemble of an {\it effective Hamiltonian} $\tilde{H}(\bm{\xi})$,
which depends on the initial condition and well approximates $H(\bm{\xi})$ as long as the system is in the metastable state,
such a metastable state is called a {\it quasi-equilibrium state} in this paper.
Now we denote the relaxation time towards a quasi-equilibrium state by $\tilde{\tau}_{\rm eq}$ and 
the relaxation time towards the true equilibrium state by $\tau_{\rm eq}$.
Then the description of quasi-equilibrium states is valid only if $\tilde{\tau}_{\rm eq}\ll t\ll \tau_{\rm eq}$,
where $t$ is the observation time.

Summarizing, we characterize the quasi-equilibrium state by the following conditions:
\begin{itemize}
\item The relaxation time towards it is much shorter than the relaxation time towards the true equilibrium state.
\item It is described by the equilibrium statistical distribution for some effective Hamiltonian $\tilde{H}(\bm{\xi})$.
This statistical distribution, which is microcanonical, canonical, or grandcanonical according to the situation, is denoted by $\tilde{\rho}_{\rm eq}(\bm{\xi})$.
The quasi-equilibrium average of a quantity $A(\bm{\xi})$ is denoted by 
$\< A\>_{\tilde{\rho}_{\rm eq}}=\sum_{\bm{\xi}\in{\cal S}^N}A(\bm{\xi})\tilde{\rho}_{\rm eq}(\bm{\xi})$.
\item $\tilde{H}(\bm{\xi})$ approximates $H(\bm{\xi})$ as long as the system is in the quasi-equilibrium state.
More specifically,
\beq
\sum_{\bm{\xi}\in{\cal S}^N}\delta\left(| H(\bm{\xi})-\tilde{H}(\bm{\xi})|>\epsilon\< |\tilde{H}|\>_{\tilde{\rho}_{\rm eq}}\right)\tilde{\rho}_{\rm eq}(\bm{\xi})<\delta
\eeq
holds for sufficiently small (but not necessarily infinitely small) positive numbers $\epsilon$ and $\delta$.
If there is a limit such that both of $\epsilon$ and $\delta$ can be made arbitrarily small, this limit is called the ``quasi-equilibrium limit''.
\end{itemize}

As a trivial example of quasi-equilibrium states, let us consider an enclosed gas placed in a thermal reservoir.
As Feynman pointed out in his textbook~\cite{Feynman_text},
such a gas ``will eventually erode its enclosure; but this erosion is a comparatively slow process, and sometime before the enclosure is appreciably eroded,
the gas will be in thermal equilibrium''.
If we consider the Hamiltonian $H$ of the gas and the enclosure, which also consists of many atoms, the true equilibrium state will be realized after the erosion.
However, if it takes a very long time for such a thing to happen,
we can consider that the erosion practically does not occur and we can replace the enclosure by just a potential wall. 
The Hamiltonian after this replacement is denoted by $\tilde{H}$,
and then the gas in the enclosure will immediately reach the ``thermal equilibrium'' described by statistical mechanics for $\tilde{H}$.
At the level of description by $H$, this ``thermal equilibrium'' is not the true equilibrium but a quasi-equilibrium state.

More nontrivial but still simple example is seen in a nearly integrable quantum system.
In an integrable system, there are many nontrivial conserved quantities $\{ I_k\}$ with $[I_k,I_l]=0$
and the Hamiltonian is written as $H=\sum_k\alpha_kI_k$.
In a nearly integrable system, a small perturbation $\eta V$ breaking the integrability is applied,
\beq
H=\sum_k\alpha_kI_k+\eta V
\eeq
with $\eta\ll 1$.
Here $[I_k,V]\neq 0$ and $\{ I_k\}$ are no longer conserved.
An equilibrium state is described by the canonical ensemble $\exp[-\beta H]/{\rm Tr}\exp[-\beta H]$~\footnote
{Relaxation in an integrable system is usually studied in an isolated system.
In that case, the use of the canonical ensemble is justified by assuming the ensemble equivalence.}.
When $\eta$ is very small, however, for relatively long time interval diverging in the limit of $\eta\rightarrow 0$,
the expectation values of $\{ I_k\}$ remain constant and do not change from the initial values $\{ I_k^{(0)}\}$.
Thus the system is first relaxed to the constrained equilibrium state, in which the expectation values of $\{ I_k\}$ are fixed to be $\{ I_k^{(0)}\}$.
This constraint is imposed by applying fictitious external fields $\{ h_k\}$ conjugate to $\{ I_k\}$, and thus the effective Hamiltonian is given by
\beq
\tilde{H}=\sum_k\alpha_kI_k-\sum_kh_k\left(I_k-I_k^{(0)}\right).
\label{eq:effective_H_integrable}
\eeq
The external fields are chosen in such a way that $I_k^{(0)}={\rm Tr}I_k\exp[-\beta\tilde{H}]/{\rm Tr}\exp[-\beta\tilde{H}]$.
The constrained equilibrium state is described by the canonical ensemble for $\tilde{H}$,
\beq
\rho_{\rm GGE}\equiv\frac{e^{-\beta\tilde{H}}}{{\rm Tr}e^{-\beta\tilde{H}}}
=\frac{\exp\left[-\sum_k\beta_kI_k\right]}{{\rm Tr}\exp\left[-\sum_k\beta_k I_k\right]}
\eeq
with $\beta_k=\beta(\alpha_k+h_k)$.
This density matrix is referred to as the ``generalized Gibbs ensemble''~\cite{Rigol2007} or
the ``pantacanonical ensemble'' in some old literatures~\cite{Truesdell1961}.
The relaxation to this kind of metastable states in an isolated quantum system is called ``prethermalization''~\cite{Berges2004,Gring2012}.

The effective Hamiltonian given by Eq.~(\ref{eq:effective_H_integrable}) satisfies the conditions on quasi-equilibrium states.
Obviously, the quasi-equilibrium limit is achieved by $\eta\rightarrow 0$.
In this way, the generalized Gibbs ensemble in an integrable system is an example of quasi-equilibrium states in our view.

\section{Model and its properties}
\label{sec:model}

\subsection{Hamiltonian}
\label{sec:Hamiltonian}
In Ref.~\cite{Mori_nonadditivity2013}, the author showed an example in which the Hamiltonian $H$ possesses the additivity but the effective one $\tilde{H}$ does not.
As a result, such a model exhibits similar properties characteristic of long-range interacting systems.
For example, the specific heat and the pseudo-magnetic susceptibility become negative in some circumstances.

Now the model studied in this paper is explained.
A microscopic state of single particle is given by $\xi_i=(\bm{q}_i,\bm{p}_i,\sigma_i)$, 
where $\bm{q}_i$ is the coordinate, $\bm{p}_i$ is the momentum conjugate to $\bm{q}_i$,
and $\sigma_i=\pm 1$ is the internal state of a particle called ``spin variable'' for simplicity.
The relevant completely additive quantity is the (pseudo-)magnetization,
\beq
M=Nm=\sum_{i=1}^N\sigma_i,
\eeq
where $m$ is the specific magnetization\footnote
{We use the specific magnetization $M/N$ instead of the magnetization density $M/V$ since it is convenient to consider the quantity per particle
when we consider quasi-equilibrium states in which the volume fluctuates.}.

The Hamiltonian is given as follows:
\beq
H(\bm{\xi})=\sum_{i=1}^N\frac{\bm{p}_i^2}{2}+\sum_{i<j}^NV_{\sigma_i,\sigma_j}(|\bm{q}_i-\bm{q}_j|)-h\sum_{i=1}^N\sigma_i.
\label{eq:Hamiltonian}
\eeq
The pair interaction potential consists of strong short-range repulsion and weak long-range attraction,
that is, we consider the Lenard-Jones-type potential.
There are three important parameters characterizing $V_{\sigma_i,\sigma_j}$:
the stable distance $R_{\sigma_i,\sigma_j}\equiv R(\sigma_i)+R(\sigma_j)$, the potential depth $V_0$, 
and the ``spring constant'' $k=\d^2V_{\sigma_i\sigma_j}(r)/\d r^2|_{r=R_{\sigma_i,\sigma_j}}$.
The stable distance depends on spin variables, and the spin degrees of freedom are coupled to the motional degrees of freedom through this dependence.

In this paper, in order to make $R_{\sigma_i,\sigma_j}$, $V_0$, and $k$ controllable independently,
we specify $V_{\sigma_i,\sigma_j}(r)$ as the following form:
\beq
V_{\sigma_i,\sigma_j}(r)=\left\{
\begin{split}
&\frac{k}{2}(R_{\sigma_i,\sigma_j}-a)^2-V_0+d_{\sigma_i,\sigma_j}\left(\frac{1}{r^{12}}-\frac{1}{a^{12}}\right) 
&\text{for } r<a_{\sigma_i,\sigma_j}, \\
&\frac{k}{2}(r-R_{\sigma_i,\sigma_j})^2-V_0 &\text{for } a_{\sigma_i,\sigma_j}\leq r\leq b_{\sigma_i,\sigma_j}, \\
&\left[\frac{k}{2}(b-R_{\sigma_i,\sigma_j})^2-V_0\right]\exp[-c_{\sigma_i,\sigma_j}(r-b)] &\text{for } r>b_{\sigma_i,\sigma_j},
\end{split}
\right.
\label{eq:potential}
\eeq
where $a_{\sigma_i,\sigma_j}$ and $b_{\sigma_i,\sigma_j}$ are chosen 
in such a way that $0<a_{\sigma_i,\sigma_j}<R_{\sigma_i,\sigma_j}<b_{\sigma_i,\sigma_j}$.
Now we impose the continuity of $\d V_{\sigma_i,\sigma_j}(r)/\d r$.
It implies
$$d_{\sigma_i,\sigma_j}=\frac{ka^{13}}{12}(R_{\sigma_i,\sigma_j}-a)$$
and
$$c_{\sigma_i,\sigma_j}=\frac{k(b-R_{\sigma_i,\sigma_j})}{V_0-\frac{k}{2}(b-R_{\sigma_i,\sigma_j})^2}.$$
In order for $c_{\sigma_i,\sigma_j}$ to be positive, it is necessary to satisfy $V_0>k(b-R_{\sigma_i,\sigma_j})^2/2$.

Throughout this paper, the parameters are set as $R(-1)=1.0$, $R(+1)=1.1$, $k=40$, 
$a_{\sigma_i,\sigma_j}=0.5R_{\sigma_i,\sigma_j}$, and $b_{\sigma_i,\sigma_j}=1.1R_{\sigma_i,\sigma_j}$.
The potential depth $V_0$ will be varied from 1.0 to 2.5.
For simplicity, we focus on the two-dimensional system.

The physical motivation of considering this model is discussed in Ref.~\cite{Mori_nonadditivity2013}.
This model is regarded as a model of some spin-crossover materials~\cite{Gutlich_review1994}.
High-spin (HS) state corresponds to $\sigma_i=+1$ and low-spin (LS) state to $\sigma_i=-1$.
The radius of a molecule is different for HS and LS states, and it is given by $R(\sigma_i)$.

We can rigorously prove that, for the system defined by Eq.~(\ref{eq:Hamiltonian}),
there exists the thermodynamic limit and the limiting entropy satisfies Eq.~(\ref{eq:entropy_additivity}).
Its proof is essentially done in a standard way given in Ruelle's textbook~\cite{Ruelle_text}.
As a result, in any genuine thermal equilibrium state, the ensemble equivalence holds and the specific heat and the magnetic susceptibility
$\chi=\d m/\d h$ are non-negative.

\subsection{Initial state and dynamics}
\label{sec:initial}

\begin{figure}[t]
\begin{center}
\includegraphics[clip,width=6cm]{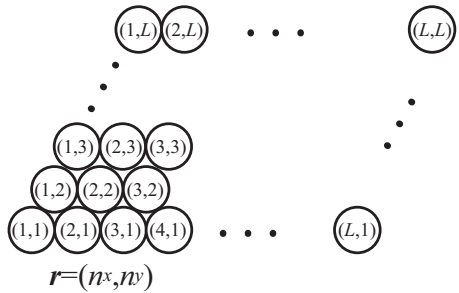}
\caption{The initial configuraiton.
. The pair of integers represents $\bm{r}_i=(n_i^{x},n_i^{y})$.}
\label{fig:initial}
\end{center}
\end{figure}

This model has relatively stable nonequilibrium configurations when $V_0$ is much greater than the temperature $T$.
These metastable configurations have the structure of the triangular lattice.
We consider such an initial condition that all the particles are regularly aligned on the triangular lattice, see Fig.~\ref{fig:initial}.
That is, the initial position of $i$th particle is given by
\beq
\bm{q}_i=(2n_i^x+n_i^y)\bm{e}_x+\sqrt{3}n_i^y\bm{e}_y,
\eeq
where $n_i^x,n_i^y=1,2,\dots, L(=N^{1/2})$, and $\bm{e}_x$ and $\bm{e}_y$ are the unit vectors directed to $x$ and $y$ axis, respectively.
All the spin variables are initially set to be $\sigma_i=-1$ (recall $R(-1)=1$).
All the momenta are initially put to be zero.
Any particle $i$ can be also labeled by the vector $\bm{r}_i=(n_i^{x},n_i^y)$.

We consider the following three situations:
\begin{itemize}
\item[(i)] The system is in contact with a thermal reservoir and the magnetization is not conserved.
The dynamics is assumed to be given by the Hamilton dynamics for $\{\bm{q}_i,\bm{p}_i\}$ and by the Monte Carlo dynamics with the Metropolis transition probabilities for $\{\sigma_i\}$.
It is expected that this dynamics leads the system to a stationary state described by the canonical ensemble without any restriction on the value of the magnetization,
which is called ``$(\beta,h)$-ensemble'' later.
\item[(ii)] The system is in contact with a thermal reservoir and the magnetization of the system is conserved.
We introduce an extra degree of freedom called the ``demon''~\cite{Creutz1983}.
The demon has a spin variable $\sigma_d$ of odd value, and in each time step,
the demon visits one of the particles at random and exchanges the spin with it.
If the demon visits the $i$th particle and $\sigma_d\sigma_i>0$, the spin does not flip.
On the other hand, if $\sigma_d\sigma_i<0$, the spin is flipped in the Metropolis transition probability.
After the flip, the spin changes as $\sigma_i\rightarrow -\sigma_i$ and demon's spin changes as $\sigma_d\rightarrow\sigma_d+2\sigma_i$.
Thus the total magnetization of the system and the demon is exactly conserved.
Technically, the demon initially has a huge spin $\sigma_d=M+N$ since we assume that all the spins are in the down state, $\sigma_i=-1$.
After a sufficiently long time, $\sigma_d$ takes the value of $+1$ or $-1$ in this algorithm.
After that, the magnetization of the system (without the demon) is almost conserved with the precision $\Delta M=\pm 1$.
Although the Monte Carlo dynamics with the conserved magnetization is more simply modeled by the Kawasaki dynamics,
the above dynamics leads the system more quickly to a stationary state described by the canonical ensemble with a fixed value of the magnetization,
which is called ``$(\beta,m)$-ensemble'' in this paper.
\item[(iii)] The system is isolated from the environment and the energy is conserved but the magnetization is not fixed.
We also consider the demon in this case, but now the demon does not have the magnetization but some amount of energy $E_d\geq 0$.
The demon visits a particle at random and try to flip the spin $\sigma_i\rightarrow -\sigma_i$.
If the energy change $\Delta E$ satisfies $E_d-\Delta E\geq 0$, this spin flip is accepted and the energy of the demon changes as $E_d\rightarrow E_d-\Delta E$.
The total energy of the system and the demon is exactly conserved.
The demon plays the role of a small thermometer, which does not change the equilibrium state of a many-body system.
This dynamics will lead the system towards a stationary state described by the microcanonical ensemble without the restriction on the value of the magnetization,
which is called ``$(\varepsilon,h)$-ensemble'' in this paper.
\end{itemize}

\begin{figure}[t]
\begin{center}
\includegraphics[clip,width=4cm]{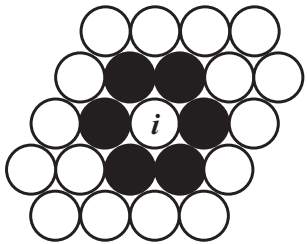}
\caption{Nearest neighbors of $i$th particle are painted in black.}
\label{fig:nearest}
\end{center}
\end{figure}

While the spin variables stochastically evolve with time according to one of the above rules,
the motional degrees of freedom evolve with time in the Hamilton equations of motion,
\begin{align}
\frac{d\bm{q}_i}{dt}&=\bm{p}_i, \\
\frac{d\bm{p}_i}{dt}&=-\sum_{j(\neq i)}\frac{\d V_{\sigma_i,\sigma_j}(|\bm{q}_i-\bm{q}_j|)}{\d\bm{q}_i}.
\end{align}
Because it is hard to solve the dynamical equations even numerically, we adopt an approximation of neglecting the interactions
except for those between the ``nearest-neighbor'' pairs on the triangular lattice.
A nearest neighbor pair $(i,j)$ is such a pair that
$(n_i^x,n_i^y)$ equals $(n_j^x\pm 1,n_j^y)$ or $(n_j^x,n_j^y\pm 1)$ or $(n_j^x-1,n_j^y+1)$ or $(n_j^x+1,n_j^y-1)$, see Fig.~\ref{fig:nearest}.
If A nearest neighbor pair $(i,j)$ is denoted by $\< i,j\>$, the Hamiltonian we consider is
\beq
H(\bm{\xi})=\sum_{i=1}^N\frac{\bm{p}_i^2}{2}+\sum_{\< i,j\>}^NV_{\sigma_i,\sigma_j}(|\bm{q}_i-\bm{q}_j|)-h\sum_{i=1}^N\sigma_i.
\label{eq:Hamiltonian_nn}
\eeq
This approximation is sufficiently accurate as long as the initial triangular lattice structure is maintained
or the particles are dilute after the breakdown of the lattice structure.

\subsection{Quasi-equilibrium states}
\label{sec:quasi}

If the time evolution starts from the initial state given in the previous section, the system first gets to a long-lived metastable state,
and after a long time, this metastable state is eventually collapsed.

Now we put the system of the shape of the triangular lattice in an infinitely extended space, and the time evolution begins~\footnote
{If we put the system in a container of volume $V$, a particle which have got out of the lattice will be reflected by a wall 
and it again approaches other particles and starts to interact with them.
In this case, the approximation of neglecting the interactions except for the ``nearest-neighbor'' ones is no longer valid
for the initial lattice structure has no meaning any more.
That is why, we consider the system put in an infinitely extended space in order to avoid this technical difficulty.
If we could solve the dynamical equations for the exact Hamiltonian~(\ref{eq:Hamiltonian}), 
it would be meaningful to consider the system in a container of finite volume.}.
Since there is no container, there is no equilibrium state.
After a very long time, the triangular lattice structure will be broken up and particles will be scattered and each particle will move freely.

\begin{figure}[t]
\begin{center}
\begin{tabular}{c}
\includegraphics[clip,width=7cm]{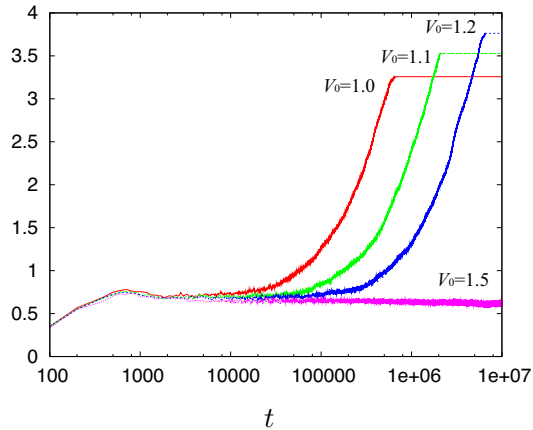} \\
$t$
\end{tabular}
\caption{Time evolution of the specific energy for various values of $V_0$.}
\label{fig:time_evolution}
\end{center}
\end{figure}

We consider the case (ii); the system is in contact with a thermal reservoir and the magnetization of the system (plus demon) is conserved.
As was explained in the previous subsection, the initial magnetizations of the system and the demon are $-N$ and $N+M$, respectively.
Figure~\ref{fig:time_evolution} shows the time evolution of the specific energy $\varepsilon=E/N$ of the system 
for $M=0$, $T=\beta^{-1}=0.26$, and various values of $V_0$.
The energy of the initial state is set to zero.
This figure shows clear two-step relaxation; first the system is trapped by a metastable state and after a long time the system leaves the metastable state.
The time evolution of the specific energy suddenly stops at some constant value, which means that the initial lattice structure is completely broken
and particles are scattered into infinitely extended space.
Each particle then behaves as a free particle and its kinetic energy is almost conserved, and hence the energy is almost conserved.
The lifetime $\tau_{\rm eq}$ of the metastability is estimated by a heuristic argument; $\tau_{\rm eq}$ will be proportional to $\exp[3\beta V_0]$
since a particle must overcome the energy barriers produced by the three neighbors in average in order to escape from the lattice.

\begin{figure}[t]
\begin{center}
\includegraphics[clip,width=7cm]{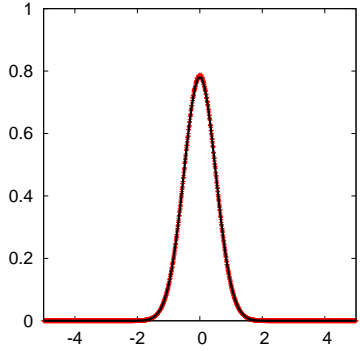}
\caption{Momentum distributions for $V_0=2.5$ calculated over the time interval $t\in[10^5,10^6]$.
The horizontal axis represents $p_x$. 
The solid line represents the Maxwell distribution.}
\label{fig:momentum}
\end{center}
\end{figure}

We measure the momentum distribution in the intermediate time scale when the system stays in metastability.
In order to numerically calculate the distribution of $p_x$,
we count the number of particles with $p_x\in [ n\Delta p_x, (n+1)\Delta p_x)$ for $n$ an integer with $|n|\leq 1000$.
The precision is $\Delta p_x=\sqrt{\< p_x^2\>}/100=\sqrt{T}/100=0.0051$.
Figure~\ref{fig:momentum} shows the momentum distribution for $V_0=2.5$ over the time interval $t\in[10^5,10^6]\ll\tau_{\rm eq}$.
It obeys the Maxwell distribution, the solid line in Fig.~\ref{fig:momentum}.

The fact that the momentum distribution in the metastable state is given by the Maxwell distribution implies that 
the metastable state is actually a quasi-equilibrium state 
with some effective Hamiltonian $\tilde{H}$ of the form
\beq
\tilde{H}(\bm{\xi})=\sum_{i=1}^N\frac{\bm{p}_i^2}{2}+\sum_{\< i,j\>}\tilde{V}_{\sigma_i,\sigma_j}(|\bm{q}_i-\bm{q}_j|)-h\sum_{i=1}^N\sigma_i,
\label{eq:eff_H}
\eeq
where $\tilde{V}_{\sigma_i,\sigma_j}$ is some effective potential.

\subsection{Effective Hamiltonian for quasi-equilibrium states}

We can easily guess the effective Hamiltonian theoretically.
We can interpret a quasi-equilibrium state as an equilibrium state with a constraint that the initial triangular lattice structure is kept.
Recall that the original interaction potential is given by Eq.~(\ref{eq:potential}), which consists of the three parts: 
$r<a_{\sigma_i,\sigma_j}$, $a_{\sigma_i,\sigma_j}\leq r\leq b_{\sigma_i,\sigma_j}$, and $r>b_{\sigma_i,\sigma_j}$.
In this paper, $a_{\sigma_i,\sigma_j}=0.5R_{\sigma_i,\sigma_j}$ and $b_{\sigma_i,\sigma_j}=1.1R_{\sigma_i,\sigma_j}$.
As long as the initial lattice structure is maintained, $|\bm{q}_i-\bm{q}_j|$ will be almost always 
between $a_{\sigma_i,\sigma_j}$ and $b_{\sigma_i,\sigma_j}$ for any ``nearest-neighbor'' pair.
Therefore, it is natural to put
\beq
\tilde{V}_{\sigma_i,\sigma_j}(r)=\frac{k}{2}(r-R_{\sigma_i,\sigma_j})^2-V_0.
\label{eq:eff_potential}
\eeq
The effective Hamiltonian for quasi equilibrium states is then given by Eq.~(\ref{eq:eff_H}) with the effective potential given by Eq.~(\ref{eq:eff_potential}),
\beq
\tilde{H}(\bm{\xi})=\sum_{i=1}^N\frac{\bm{p}_i^2}{2}+\sum_{\< i,j\>}\left[\frac{k}{2}(|\bm{q}_i-\bm{q}_j|-R_{\sigma_i,\sigma_j})^2-V_0\right]
-h\sum_{i=1}^N\sigma_i.
\label{eq:elastic_spin}
\eeq
A variant of this Hamiltonian has been first introduced in the study of spin-crossover materials in Ref.~\cite{Nishino2007},
and the connection with more fundamental models such as Eq.~(\ref{eq:Hamiltonian}) was first discussed in Ref.~\cite{Mori_nonadditivity2013},
in which it was found out that the model of the Hamiltonian~(\ref{eq:elastic_spin}) does not possess additivity.
Because of this fact, this model would be interesting not only in the context of the study of spin-crossover materials 
but also in the context of the study of fundamental problems in statistical mechanics.
The model with the Hamiltonian~(\ref{eq:elastic_spin}) is refereed to as the ``elastic-spin model''~\cite{Mori_nonadditivity2013}.
We can say in this way: equilibrium statistical mechanics of the elastic-spin model describes 
quasi-equilibrium states of the particle system with the Hamiltonian~(\ref{eq:Hamiltonian}) for initial conditions of triangular lattice structure.

\section{Properties of quasi-equilibrium states}
\label{sec:properties}

Let us investigate the properties of quasi-equilibrium states of the model~(\ref{eq:Hamiltonian}),
that is, the {\it equilibrium} properties of the elastic spin model~(\ref{eq:elastic_spin}).
As was mentioned in Sec.~\ref{sec:initial}, we consider the following three cases,
(i) $(\beta,h)$-ensemble: the case where the system is in contact with a thermal reservoir and the magnetization is not conserved,
(ii) $(\beta,m)$-ensemble: the case where the system is in contact with a thermal reservoir and the magnetization is conserved,
and (iii) $(\varepsilon,h)$-ensemble: the case where the system is isolated from the environment and the energy is conserved but the magnetization is not conserved.

Since we have derived the effective Hamiltonian in a heuristic way,
one might wonder whether quasi-equilibrium states of Eq.~(\ref{eq:Hamiltonian_nn}) are really described 
by the equilibrium statistical mechanics of $\tilde{H}$ given by Eq.~(\ref{eq:elastic_spin}).
Therefore, we compare the former with the latter in all the three cases.
That is, we compare time averages of some quantities over the metastable time interval calculated for Eq.~(\ref{eq:Hamiltonian_nn}) 
with the corresponding equilibrium ensemble averages for Eq.~(\ref{eq:elastic_spin}).

\subsection{Nonadditivity}
\label{sec:nonadditivity}

Before going on to the investigation of the three cases, we first show that the elastic spin model is {\it not} additive.
In Sec.~\ref{sec:additivity}, we have shown that the system is nonadditive 
if we can divide the system into the two subsystems $A$ and $B$ without a macroscopic amount of work.
Especially, if Eq.~(\ref{eq:additivity_can}) does not hold, 
Eqs.~(\ref{eq:additivity_micro}) and (\ref{eq:entropy_additivity}) also do not hold and the system is thus not additive.
Therefore, in order to show that the elastic spin model is {\it not} additive, it is sufficient to consider the canonical ensemble with both the magnetizations of $A$ and $B$ held fixed.
When the magnetization is conserved, the homogeneous magnetic field $h$ does not play any role, and hence we put $h=0$ in this subsection.
The constant energy $-V_0$ in the effective potential is also not important, so we put $V_0=0$.

We regard every spin $i$ with $n_i^x\leq L/2$ as a part of the subsystem $A$ and otherwise as a part of the subsystem $B$.
We decompose the effective Hamiltonian as
\beq
\tilde{H}(\bm{\xi})=H_A(\bm{\xi}_A)+H_B(\bm{\xi}_B)+\eta H_{\rm int}(\bm{\xi}_A,\bm{\xi}_B)
\eeq
where
\begin{align}
H_A&=\sum_{i\in A}\frac{\bm{p}_i^2}{2}
+\sum_{\substack{\< i,j\> \\ i,j\in A}}\frac{k}{2}(|\bm{q}_i-\bm{q}_j|-R_{\sigma_i,\sigma_j})^2 \\
H_B&=\sum_{i\in B}\frac{\bm{p}_i^2}{2}
+\sum_{\substack{\< i,j\> \\ i,j\in B}}\frac{k}{2}(|\bm{q}_i-\bm{q}_j|-R_{\sigma_i,\sigma_j})^2 \\
H_{\rm int}&=\sum_{\substack{\< i,j\> \\ i\in A,j\in B}}\frac{k}{2}(|\bm{q}_i-\bm{q}_j|-R_{\sigma_i,\sigma_j})^2
\end{align}
The work parameter $\eta$ is first set to $\eta=1$ and is turned off very slowly.
We consider the case in which $M_A=N/2$ ($m_A=1$) and $M_B=-N/2$ ($m_B=-1/2$).
Both $M_A$ and $M_B$ should be conserved during the quasi-static isothermal process in order to investigate the additivity,
we prepare the two ``demons'' each for the subsystem $A$ and the subsystem $B$.
The demons for the subsystem $A$ and $B$ exchange the magnetization only with the subsystem $A$ and $B$, respectively.
Except for the existence of the two demons, the dynamics is the same as that for the case (ii) in Sec.~\ref{sec:initial}.

\begin{figure}[t]
\begin{center}
\includegraphics[clip,width=7cm]{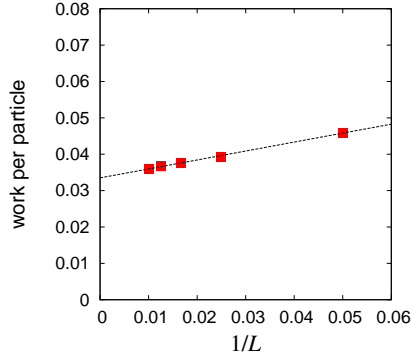}
\caption{The work per particle done by the system during the quasi-static isothermal process dividing the system into the subsystems $A$ and $B$
with $m_A=1$ and $m_B=-1$.
The horizontal axis is the inverse of the system size, $1/L=1/N^{1/2}$.
The dashed line is the linear fitting of the data.}
\label{fig:additivity}
\end{center}
\end{figure}

Figure~\ref{fig:additivity} shows the work done by the system during the quasi-static isothermal process from $\eta=1$ to $\eta=0$.
As a function of the system size, the work per particle does {\it not} vanish in the thermodynamic limit;
Eq.~(\ref{eq:additivity_can}) does not hold and thus {\it the elastic spin model is not additive}.

It should be noted that the total energy is proportional to the system size, that is, the {\it extensivity} still holds~\footnote
{In a usual long-range interacting system, the extensivity is restored by Kac's prescription, 
which is a theoretical procedure to extract nontrivial macroscopic properties from the Hamiltonian.
However, in the elastic spin model, such a theoretical prescription by hand is unnecessary; The extensivity is automatically satisfied.
Remarkably, the elastic spin model is, nevertheless, nonadditive.}.

The reason why the elastic spin model can be extensive but nonadditive is understood by the following consideration.
In the elastic spin model, any particle only interacts with the nearest neighbors.
In this sense, the elastic spin model is local and the energy is proportional to the system size; the extensivity is satisfied.
On the other hand, the effective potential between the nearest neighbors is highly nonlocal in the sense that the interparticle force does not vanish in long distances.
This nonlocality stems from the restriction that the initial triangular lattice structure is kept.
The nonadditivity of the elastic spin model does not contradict the rigorous results of statistical mechanics, 
which tell us that any macroscopic system is additive {\it as long as the interparticle potential is local}~\footnote
{Strictly speaking, the term ``local'' here means the tempering and the stability conditions on the interaction potential, see Ref.~\cite{Ruelle_text}.}.

Since the effective Hamiltonian is valid only for $t\lesssim\tau_{\rm eq}\propto\exp[3\beta V_0]$,
the system will restore the additivity by breaking the lattice structure spontaneously.
The nonadditivity appears while this ultimately unstable lattice structure is kept.
Thus {\it the nonadditivity can emerge through metastable configurational structure even in a short-range interacting system}.
Since the true equilibrium state should be additive, this configurational structure should be eventually broken.

\subsection{Case (i): $(\beta,h)$-ensemble}
\label{sec:case(i)}

Now we compare the time averages of some quantities in the case (i) over the quasi-equilibrium time interval $\tilde{\tau}_{\rm eq}\ll t\ll\tau_{\rm eq}$
with the corresponding ensemble average calculated by the $(\beta,h)$-ensemble with $\tilde{H}$.
The ensemble average can be calculated by the Monte Carlo sampling, 
that is, the time average under the same dynamics as that of the case (i) in Sec.~\ref{sec:initial} for $\tilde{H}$ instead of $H$.

\begin{figure}[tb]
\begin{center}
\begin{tabular}{cc}
\includegraphics[clip,width=6cm]{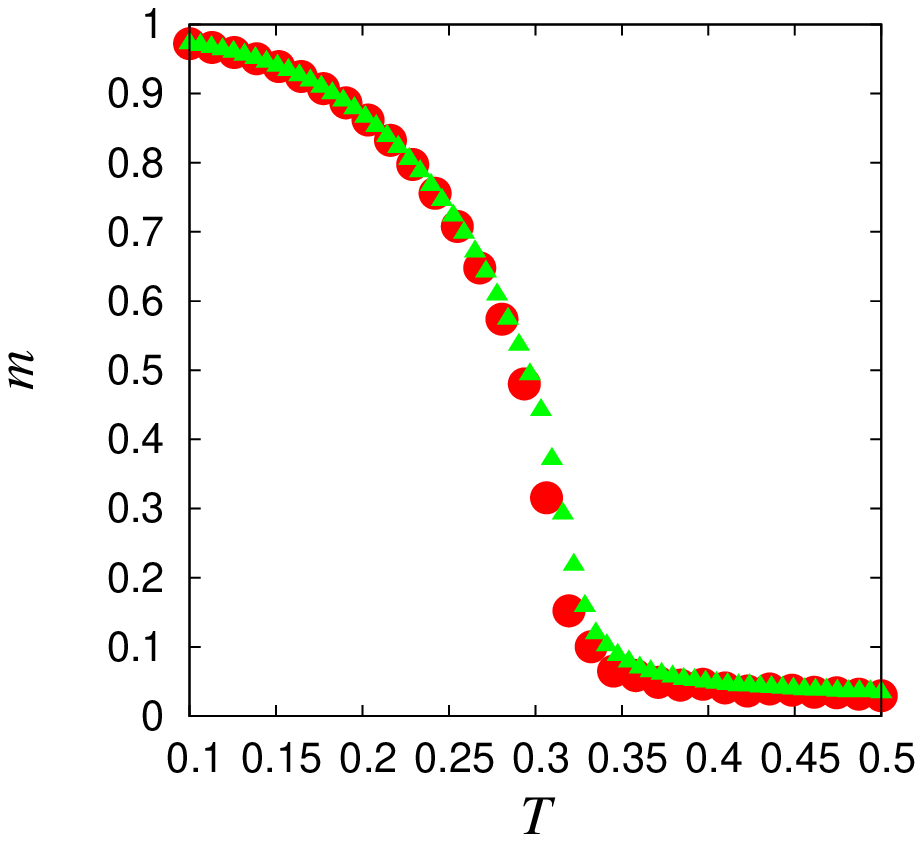}&
\includegraphics[clip,width=6cm]{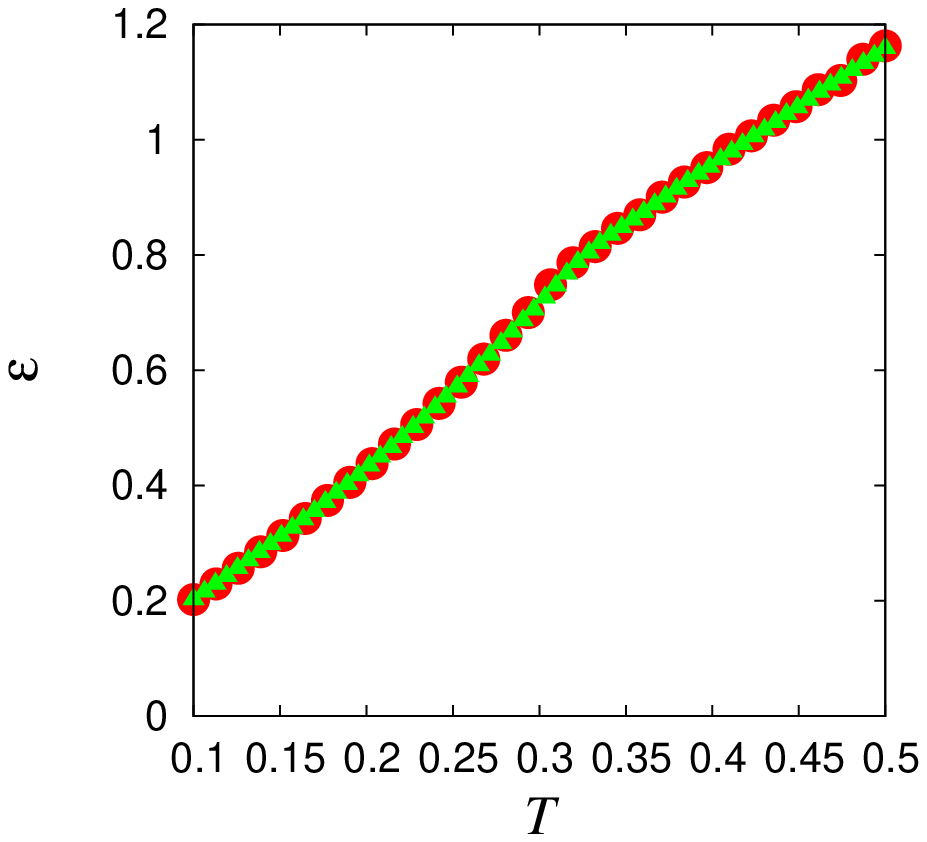}
\end{tabular}
\caption{(Left) The magnetization as a function of the temperature, (Right) the specific energy as a function of the temperature.
The circles correspond to the quasi-equilibrium time averages under the dynamics of the case (i) in Sec.~\ref{sec:initial} for $H$.
The triangles correspond to the $(\beta,h)$-ensemble averages with $\tilde{H}$.}
\label{fig:comparison_TH}
\end{center}
\end{figure}

Figure~\ref{fig:comparison_TH} show the curves of the magnetization $m=\sum_{i=1}^N\<\sigma_i\>/N$ 
and the specific energy $\varepsilon=\< H\>/N$, respectively,
as a function of the temperature for zero magnetic field $h=0$.
Brackets mean the time average over $t\in[10^5,10^6]$ for circles and the $(\beta,h)$-ensemble average for $\tilde{H}$ for triangles.
These figures show that the time average in quasi-equilibrium states agrees very well to the $(\beta,h)$-ensemble average for $\tilde{H}$.
Thus it has been confirmed that quasi-equilibrium states indeed can be described by the canonical ensemble with the effective Hamiltonian.

\subsection{Case (ii): $(\beta,m)$-ensemble}
\label{sec:case(ii)}

We shall compare the time averages of some quantities under the dynamics given as the case (ii) in Sec.~\ref{sec:initial} 
over the quasi-equilibrium time interval $t\in[10^5,10^6]$ to the corresponding $(\beta,m)$-ensemble averages with $\tilde{H}$.

\begin{figure}[t]
\begin{center}
\begin{tabular}{cc}
\includegraphics[clip,width=7cm]{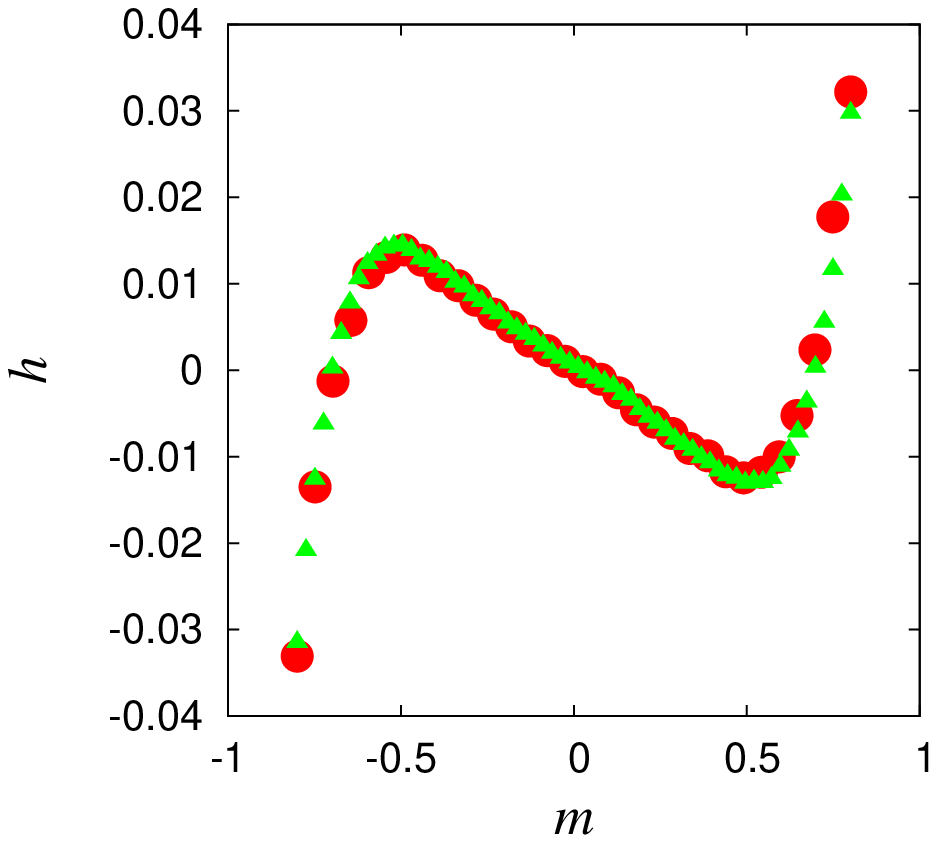}&
\includegraphics[clip,width=7cm]{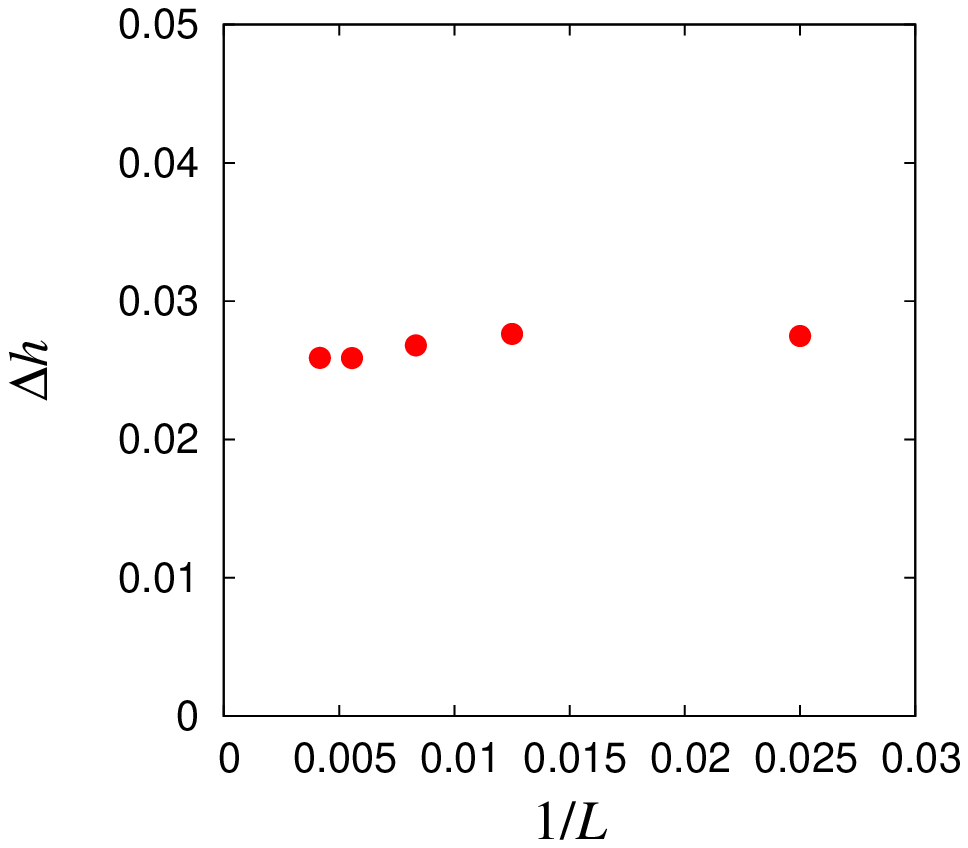}
\end{tabular}
\caption{(Left) The magnetic field as a function of the magnetization for $T=0.26$.
The circles correspond to the quasi-equilibrium time averages under the dynamics of the case (ii) in Sec.~\ref{sec:initial} for $H$.
The triangles correspond to the $(\beta,m)$-ensemble averages with respect to $\tilde{H}$.
(Right) The system-size dependence of $\Delta h=h_h-h_l$, where $h_h$ and $h_l$ are the local maximum and the local minimum values of $h$, respectively.}
\label{fig:comparison_TM}
\end{center}
\end{figure}

The left of Fig.~\ref{fig:comparison_TM} shows the magnetic field $h$ as a function of $m$.
Here, the magnetic field $h$ is defined as $h=\d f(\beta,m)/\d m$.
Since the total amount of the magnetization of the system and the demon is conserved,
the average magnetization of the demon is related to $h$ as
\beq
h=\frac{1}{2\beta}\ln\frac{1+\<\sigma_d\>}{1-\<\sigma_d\>},
\label{eq:h_demon}
\eeq
see Ref.~\cite{Mori_nonadditivity2013}.
Therefore, it is not necessary to compute the partial derivative of the free energy directly.
Instead, we can obtain $h$ by measuring $\<\sigma_d\>$ with the help of Eq.~(\ref{eq:h_demon}).

As is clearly observed in Fig.~\ref{fig:comparison_TM}, the time averages over quasi-equilibrium time interval excellently agree with the $(\beta,m)$-ensemble averages associated with $\tilde{H}$.

In Fig.~\ref{fig:comparison_TM}, we can also see the negative susceptibilities.
The susceptibility is defined as
$\chi=\d m/\d h=(\d h/\d m)^{-1}$.
In the $(\beta,h)$-ensemble, $\chi$ is always nonnegative because $\chi$ is also expressed as
\beq
\chi=\frac{\beta}{N}(\<M^2\>-\<M\>^2).
\eeq
Thus negative values of $\chi$ in the $(\beta,m)$-ensemble is an evidence of ensemble inequivalence in quasi-equilibrium states.

In short-range interacting systems, the negative slope of the magnetic field can be observed for finite systems,
but $\Delta h$ vanishes in the thermodynamic limit.
In order to confirm that the observed negative susceptibilities are {\it not} due to finite size effect,
we have also examined the system size dependence of the difference between the local maximum of the magnetic field, $h_h$ 
and the local minimum of the magnetic field, $h_l$ associated with the negative magnetic susceptibilities.
In the right of Fig.~\ref{fig:comparison_TM}, we find that $\Delta h=h_h-h_l$ does not shrink as the system size increases.

\subsection{Case (iii): $(\varepsilon,h)$-ensemble}
\label{sec:case(iii)}

Finally, let us consider the case (iii).
In the case (iii), interesting things happen when we introduce the $g$-fold degeneracies to the spin up state~\footnote
{The introduction of $g$ slightly modifies the dynamics of spins.
We multiply the factor $g/(g+1)$ and $1/(g+1)$ to the probability of flipping a spin 
from $\sigma_i=-1$ to $+1$ and from $\sigma_i=+1$ to $-1$, respectively.}, that is,
\beq
\sigma_i\in\{\underbrace{+1,+1,\dots,+1}_{g},-1\}.
\eeq
This kind of degeneracy is actually important in spin-crossover compounds~\cite{Gutlich_review1994}, 
which have motivated the study of the elastic-spin model~\cite{Nishino2007}.
In this subsection, we set $g=20$ and $h=-0.15$.

\begin{figure}[t]
\begin{center}
\begin{tabular}{cc}
(a)&(b)\\
\includegraphics[clip,width=7cm]{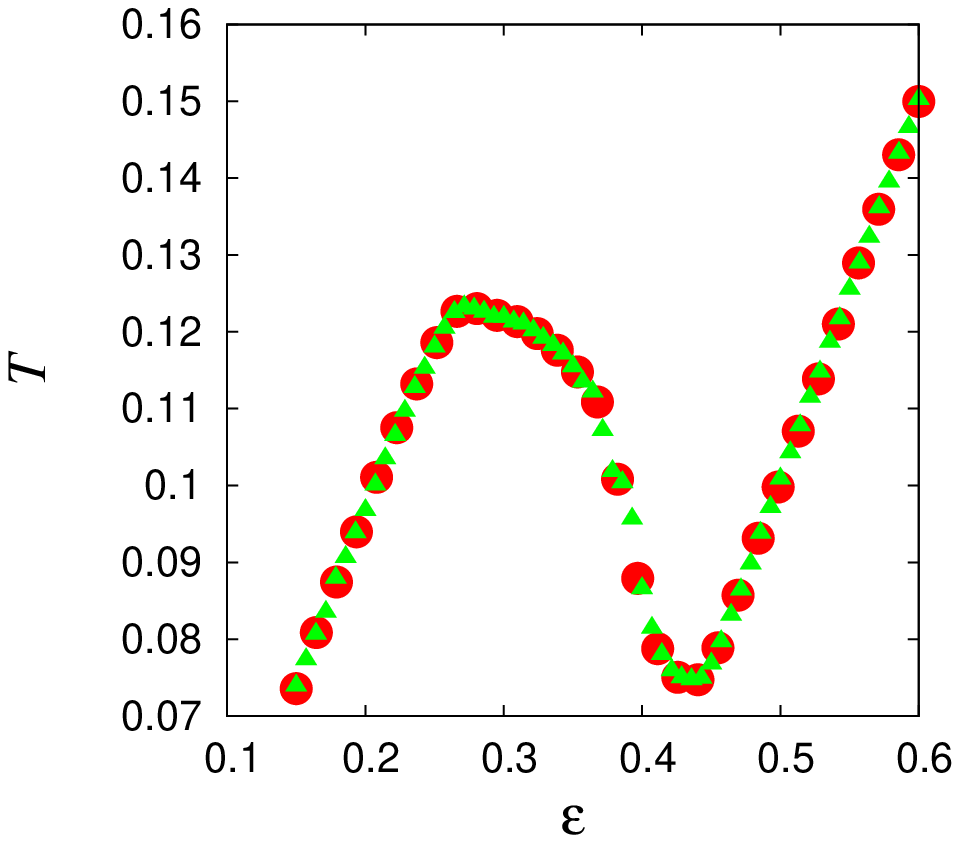}&
\includegraphics[clip,width=7cm]{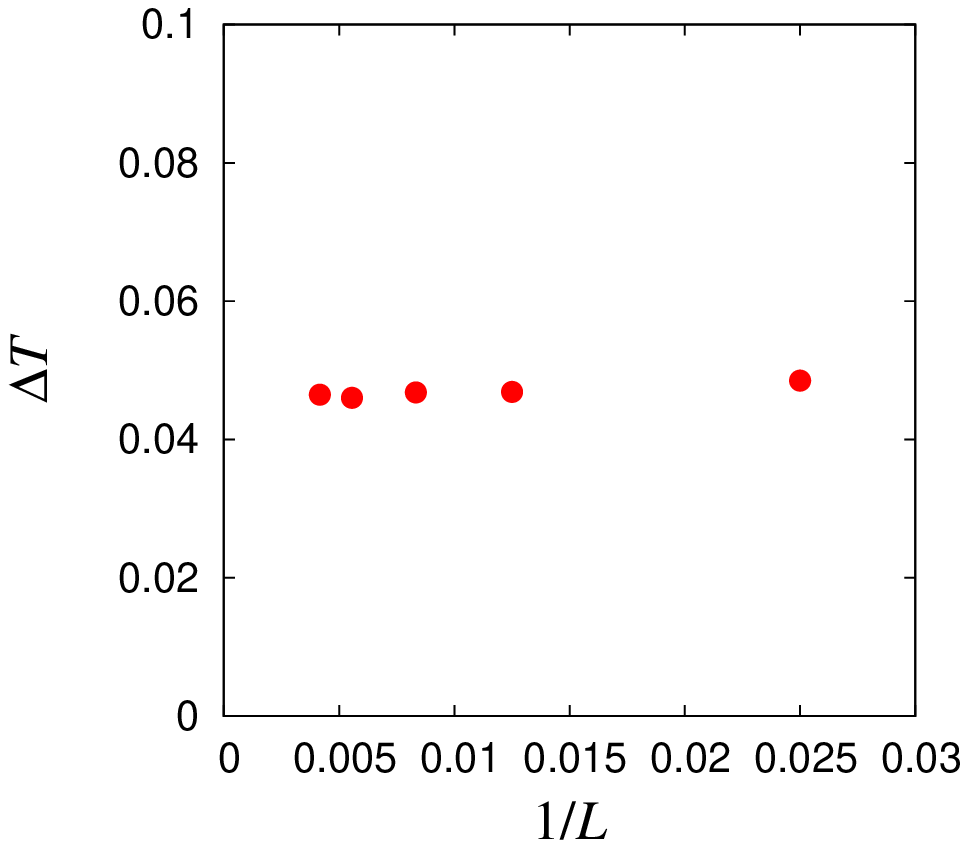}
\end{tabular}
\caption{(a) The temperature as a function of the specific energy.
The degeneracy of the up-spin state is $g=20$ and the magnetic field is set to $h=-0.15$.
The circles correspond to the quasi-equilibrium time averages under the dynamics of the case (ii) in Sec.~\ref{sec:initial} for $H$.
The triangles correspond to the $(\beta,m)$-ensemble averages with $\tilde{H}$.
(b) The system-size dependence of $\Delta T=T_h-T_l$, where $T_h$ and $T_l$ are the local maximum and the local minimum values of $T$, respectively.}
\label{fig:comparison_EH}
\end{center}
\end{figure}

In Fig.~\ref{fig:comparison_EH} (a), we show the energy dependence of the temperature.
In an isolated system, the temperature is defined as $T=\beta^{-1}=(\d s/\d\varepsilon)^{-1}$.
Since the demon exchanges the energy with the system, the equilibrium state of the demon is described by the canonical ensemble.
Since demon's energy $\varepsilon_d$ takes any positive real number, the average value of $\varepsilon_d$ is given by
\beq
\<\varepsilon_d\>=\frac{\int_0^{\infty}\varepsilon e^{-\beta\varepsilon}d\varepsilon}{\int_0^{\infty}e^{-\beta\varepsilon}d\varepsilon}
=\frac{1}{\beta}=T.
\eeq
Therefore, the temperature of the system is obtained by measuring the average value of demon's energy.
The demon plays the role of a thermometer.

In Fig.~\ref{fig:comparison_EH} (a), we can clearly observe the negative specific heats.
The specific heat is defined as $c=\d\varepsilon/\d T=(\d T/\d\varepsilon)^{-1}$.
In the canonical ensemble, it is also expressed as
\beq
c=\frac{\beta^2}{N}\left(\<E^2\>-\<E\>^2\right),
\eeq
where $E$ is the energy of the system.
This is always nonnegative.
Thus, the negative value of the specific heat is also a clear evidence of the ensemble inequivalence in quasi-equilibrium states.

We also examine the system size dependence of the difference between the local maximum temperature $T_h$ and the local minimum temperature $T_l$
associated with the negative specific heats in order to see that the negative specific heats are not due to finite size effect.
Figure.~\ref{fig:comparison_EH} (b) clearly shows that $\Delta T=T_h-T_l$ does not vanish in large system sizes in contrast to usual short-range interacting systems.

\section{Effective long-range spin-spin interactions}
\label{sec:effective}

Nonadditivity is a characteristic of long-range interacting systems.
Therefore, the nonadditivity of the elastic spin model implies that some effective long-range spin-spin interactions emerge in this model.

In Ref.~\cite{Mori_nonadditivity2013}, it was clarified that the effective spin-spin interaction mediated by the elastic motion by $\{\bm{q}_i\}$ and $\{\bm{p}_i\}$ has the interaction range proportional to the system size.
The effective spin-spin interactions are obtained by eliminating the degrees of freedom of $\{\bm{q}_i\}$ and $\{\bm{p}_i\}$ from the elastic spin model.
We here assume that the effective spin Hamiltonian is written as
\beq
\tilde{H}_{\rm spin}=\sum_{i<j}J_{ij}\sigma_i\sigma_j.
\label{eq:eff_spin}
\eeq
Here we put $h=0$ for simplicity.

Although we consider the triangular lattice, the position of each site $i$ is labeled by $\bm{r}_i=(n_i^x,n_i^y)$ as in Fig.~\ref{fig:initial}.
The region of the system $\Gamma$ is identified in such a way that each lattice point $(n_i^x,n_i^y)$ in $\Gamma$ corresponds to each spin with $\bm{r}_i=(n_i^x,n_i^y)$.
The number of spins $N$ equals the number of lattice points belonging to $\Gamma$.
Now we consider some fixed $\gamma\subset\mathbb{R}^d$ (in our case $d=2$) with the unit volume and the system is made large as $\Gamma=L\gamma$, remember the setup given in Sec.~\ref{sec:setup}.

In the effective spin Hamiltonian, the microscopic state is just given by $\xi_i=\sigma_i$
because $\bm{q}_i$ and $\bm{p}_i$ have disappeared from the description.
We use the same notation like ${\cal S}$ or $\bm{\xi}$ for the effective spin Hamitonian.

Since it is very hard to derive the effective spin-spin interactions $\{J_{ij}\}$ analytically,
in Ref.~\cite{Mori_nonadditivity2013} the author estimated $J_{ij}$ from the numerical data of $\{\<\sigma_i\>\}$ and $\{C_{ij}=\<\sigma_i\sigma_j\>\}$.
Now we consider the disordered phase, $\<\sigma_i\>=0$ for all $i$.
The estimation of $J_{ij}$ from $C_{ij}$ is done by using the formula
\beq
\beta J_{ij}=\delta_{ij}-\left(C^{-1}\right)_{ij},
\label{eq:estimation}
\eeq
where $C^{-1}$ is the inverse matrix of $(C_{ij})$.
This formula is not exact in general, but it becomes exact when $J_{ij}$ takes the form
\beq
J_{ij}=L^{-d}\phi\left(\frac{\bm{r}_i}{L},\frac{\bm{r}_j}{L}\right)
\label{eq:nonadditive}
\eeq
with some function $\phi$ satisfying
\beq
\int_{\gamma}d^d\bm{x}\int_{\gamma}d^d\bm{y}\phi(\bm{x},\bm{y})<+\infty.
\label{eq:integrability}
\eeq
We shall call the scaling of Eq.~(\ref{eq:nonadditive}) the {\it nonadditive scaling}~\footnote
{If we consider the power-law potential, $\phi(\bm{x},\bm{y})\sim 1/|\bm{x}-\bm{y}|^{\alpha}$,
the condition~(\ref{eq:integrability}) implies $\alpha<d$.
As is known, $\alpha<d$ belongs to the nonadditive regime~\cite{Campa_review2009,Levin_review2014}.}.
In Ref.~\cite{Mori_nonadditivity2013}, it was clarified that the effective spin-spin interactions actually satisfy the nonadditive scaling at least for $d=2$ and $\gamma=[0,1]^2$.
Therefore, the validity of the formula~(\ref{eq:estimation}) for the elastic spin model was confirmed ex posto facto.

For a while, let us leave the elastic spin model and consider the general spin systems with Eqs.~(\ref{eq:eff_spin}) and (\ref{eq:nonadditive}).
The nonadditive scaling means that the interaction is very weak and proportional to the inverse of the volume of the system, but its range is very long and proportional to the system size $L$.
It is rigorously shown that the thermodynamic limit of the entropy density exists in general spin systems given by Eqs.~(\ref{eq:eff_spin}) and (\ref{eq:nonadditive}), but the system is nonadditive.
As a result, the entropy density $s_{\gamma}(\varepsilon,m)$ depends on $\gamma$ and is not necessarily concave in contrast to the case of additive systems, see Sec.~\ref{sec:additivity}.

In the periodic boundary conditions, because of the translational symmetry, 
we can write $\phi(\bm{x},\bm{y})=\phi(\bm{x}-\bm{y})$.
In that case, Eq.~(\ref{eq:nonadditive}) means 
\beq
J_{ij}=L^{-d}\phi\left(\frac{\bm{r}_i-\bm{r}_j}{L}\right).
\label{eq:Kac}
\eeq
The above form corresponds to $\gamma=L^{-1}$ in the Kac scaling, $J_{ij}=\gamma^d\phi(\gamma(\bm{r}_i-\bm{r}_j))$~\cite{Kac1963}.
In the periodic boundary condition, the {\it exactness of the mean-field theory} and its violation can be rigorously derived~\cite{Mori_analysis2010,Mori_instability2011,Mori_microcanonical2012,Mori_equilibrium2012}.
The exactness of the mean-field theory means that the thermodynamic function is independent of the precise form of the function $\phi$ under the normalization $\int_{\gamma}d^d\bm{x}\phi(\bm{x})=1$, and thus the thermodynamic function is identical to that of the mean-field model in which $\phi(\bm{x})=1$.
See Ref.~\cite{Mori_statphys2013} for the detail on this subject.
That is why, in a wide parameter region called the ``MF region'', which is specified by the ``region A'' and a part of the ``region B'' in Ref.~\cite{Mori_statphys2013}, equilibrium states of the elastic spin model can be essentially understood by examining the much simpler mean-field model,
\beq
H_{\rm MF}=-\frac{1}{2N}\sum_{ij}\sigma_i\sigma_j.
\eeq 
This is the great simplification of the problem.
On the other hand, in the parameter region called the ``non-MF region'', which is specified by the other part of the ``region B'' and the ``region C'' in Ref.~\cite{Mori_statphys2013}, the strong non-mean-field behavior like the macroscopic inhomogeneity is observed~\cite{Mori_instability2011}.
In this case, the analysis of the mean-field model does not help us to understand the equilibrium properties of the spin system.

Now let us go back to the elastic spin model.
In the previous work~\cite{Mori_nonadditivity2013}, the form of $\phi(\bm{x},\bm{x}_c)$ for $\gamma=[0,1]^2$ was displayed by using the formula~(\ref{eq:estimation}) and it was found that it is independent of the temperature or the specific ensemble.
Here $\bm{x}_c$ denotes the central position of the system.
Moreover, the previous work has shown that the approximation 
$\phi(\bm{x},\bm{y})\simeq\phi(\bm{x}-\bm{y},\bm{x}_c)\equiv\phi(\bm{x}-\bm{y})$ is valid by comparing the equilibrium states of the elastic spin model with those of the spin model~(\ref{eq:eff_spin}) with $J_{ij}=L^{-d}\phi((\bm{r}_i-\bm{r}_j)/L,\bm{x}_c)$.
It means that we can use Eq.~(\ref{eq:Kac}) even if the boundary condition is not periodic and Eq.~(\ref{eq:Kac}) is not exactly satisfied.

\begin{figure}[t]
\begin{center}
\begin{tabular}{cc}
\includegraphics[clip,width=7.5cm]{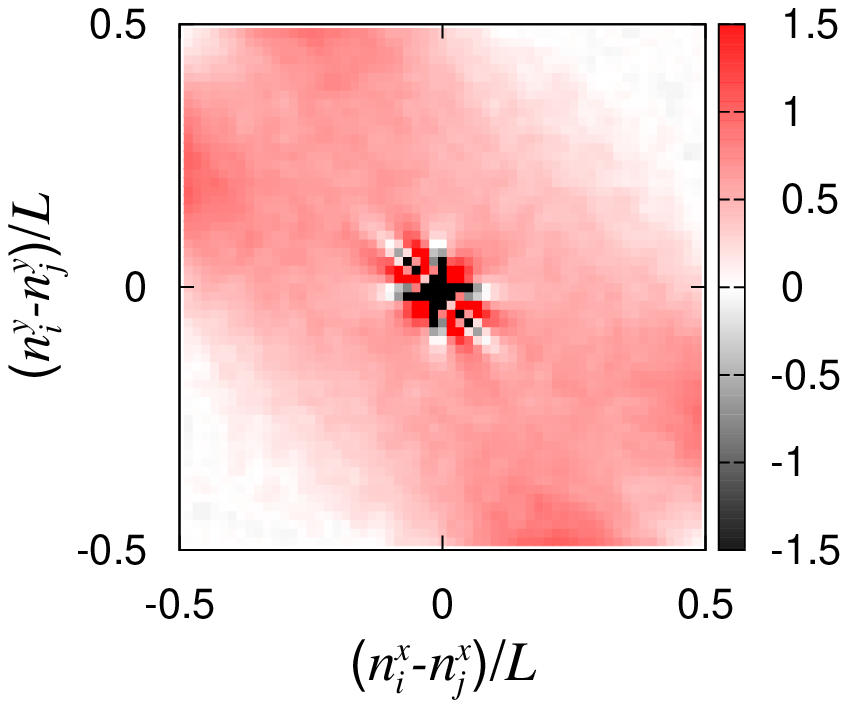}&
\includegraphics[clip,width=7.5cm]{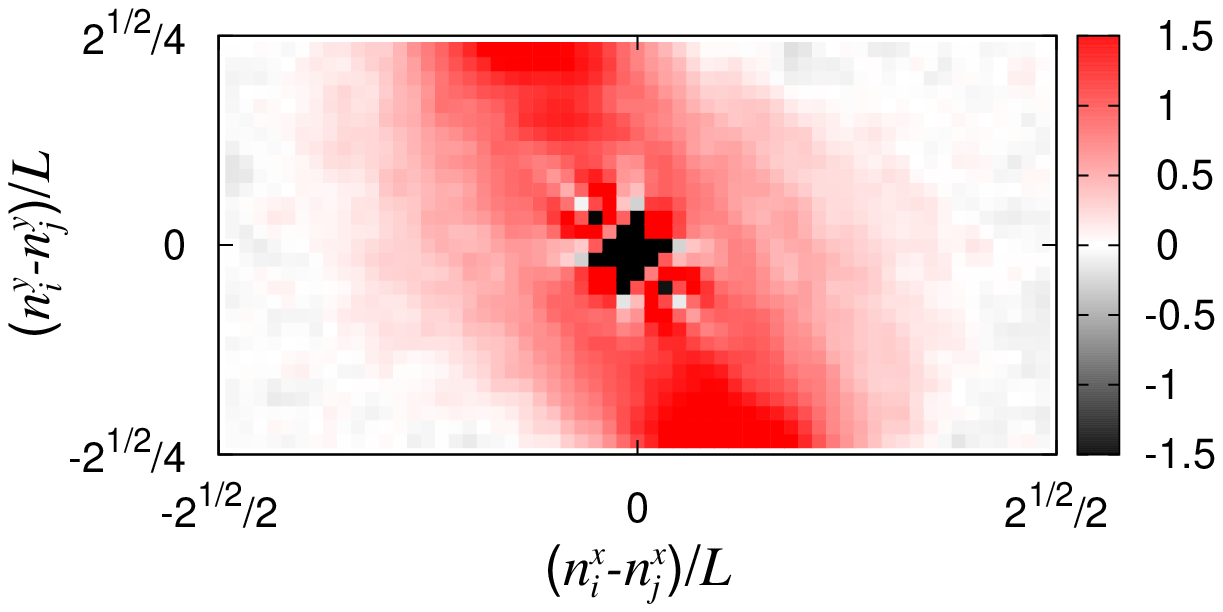}
\end{tabular}
\caption{The details of $\phi(\bm{r}_i/L,\bm{r}_j/L)$ for (Left) $\gamma=[0,1]^2$ and (Right) $\gamma=[0,\sqrt{2}]\times[0,1/\sqrt{2}]$.
The site $j$ is fixed at the center, $\bm{r}_j/L=\bm{x}_c$.
Clearly, the function $\phi(\bm{x})$ strongly depends on the shape of the system $\gamma$.}
\label{fig:map_int}
\end{center}
\end{figure}

Now we shall show the dependence of the scaled potential $\phi(\bm{x},\bm{y})$ on $\gamma$, the shape of the system.
In Fig.~\ref{fig:map_int}, $\phi(\bm{x},\bm{x}_c)$ for (a) $\gamma=[0,1]^2$ and (b) $\gamma=[0,\sqrt{2}]\times[0,1/\sqrt{2}]$ are demonstrated.
We can see that the form of $\phi(\bm{x},\bm{x}_c)$ strongly depends on $\gamma$.

\section{Conclusion}
\label{sec:conclusion}

In this paper, the definition of additivity and its consequence were discussed,
and then it was shown that a short-range interacting system can violate the additivity
when it is in a quasi-equilibrium state.
Some rigorous results of statistical mechanics~\cite{Ruelle_text} tell us that any short-range interacting system with some natural properties, that is, the {\it stability} and the {\it tempering}, is additive.
It excludes the possibility that the influence of short-range interactions spreads out over long distances
and the system becomes nonadditive in an equilibrium state.
From this point of view, the result obtained in this work implies that the natural conditions, the stability and the tempering, do not necessarily hold for the effective Hamiltonian describing the quasi-equilibrium states
even if they hold for the ``genuine'' Hamiltonian.

In our model, the elastic motion mediates the long-range spin-spin couplings in quasi-equilibrium states.
This long-range coupling stems from the size difference between the particle with $\sigma_i=+1$ and the particle with $\sigma_i=-1$.
Although we call $\sigma_i$ the ``spin variable'', it represents the internal state of the particle and
the ``magnetic'' phase transition in our model corresponds to a kind of structural phase transitions.
Therefore, it is expected that the {\it quasi-equilibrium nonadditivity} will be found more broadly in systems exhibiting structural phase transitions.
It is a future problem to understand the possible relation between the structural phase transitions and the nonadditivity.

Long-range interacting systems also show some interesting dynamical properties~\cite{Campa_review2009,Levin_review2014}
such as the existence of the quasi-stationary states~\cite{Antoni-Ruffo1995,Yamaguchi2004,Baldovin2009,Gupta-Mukamel2010,Kastner2011}
 (they are different from quasi-equilibrium states discussed in this paper).
To investigate the dynamical properties of the model given by Eq.~(\ref{eq:Hamiltonian}) would be also interesting.
In this model, the effective long-range interactions emerge as a result of the transmission of the influence of the short-range forces.
Therefore, in short timescales the system will look like a short-range interacting system,
but in long timescales it will behave like a long-range interacting system.
This feature might bring some interesting dynamical characteristics.
This issue will be studied elsewhere.

\subsection*{Acknowledgement}
The author is grateful to Seiji Miyashita for the discussions and warm encouragement.
The author also would like to thank Stefano Ruffo and Hugo Touchette for suggestive comments and discussions.
This work was supported by the Sumitomo Foundation (grant No. 120753).
The computation in this work has been partially done using the facilities of the Supercomputer Center, Institute for Solid State Physics, the University of Tokyo.

\end{document}